# Deterministic generation of single soliton Kerr frequency comb in microresonators by a single shot pulsed trigger

ZHE KANG,[1,7] FENG LI,[1,2,7,8] JINHUI YUAN,[2,3,7,9] K. NAKKEERAN,[4] J. NATHAN KUTZ,[5] QIANG WU,[6] CHONGXIU YU,[3] P. K. A. WAI,[1,2]

[1] *Photonics Research Centre, Department of Electronic and Information Engineering, The Hong Kong Polytechnic University, Hung Hom, Hong Kong SAR.*
[2] *Hong Kong Polytechnic University Shenzhen Research Institute, 518057, Shenzhen, China.*
[3] *State Key Laboratory of Information Photonics and Optical Communications, Beijing University of Posts and Telecommunications, 100876, Beijing, China.*
[4] *School of Engineering, University of Aberdeen, Aberdeen, AB24 3UE, United Kingdom.*
[5] *Department of Mathematics, University of Washington, Seattle, WA 98195, United States.*
[6] *Department of Physics and Electrical Engineering, Northumbria University, Newcastle upon Tyne, NE1 8ST, United Kingdom.*
[7] *These authors contributed equally*
[8] *lifeng.hk@gmail.com*
[9] *yuanjinhui81@163.com*

**Kerr soliton frequency comb generation in monolithic microresonators recently attracted great interests as it enables chip-scale few-cycle pulse generation at microwave rates with smooth octave-spanning spectra for self-referencing. Such versatile platform finds significant applications in dual-comb spectroscopy, low-noise optical frequency synthesis, coherent communication systems, etc. However, it still remains challenging to straightforwardly and deterministically generate and sustain the single-soliton state in microresonators. In this paper, we propose and theoretically demonstrate the excitation of single-soliton Kerr frequency comb by seeding the continuous-wave driven nonlinear microcavity with a pulsed trigger. Unlike the mostly adopted frequency tuning scheme reported so far, we show that an energetic single shot pulse can trigger the single-soliton state deterministically without experiencing any unstable or chaotic states. Neither the pump frequency nor the cavity resonance is required to be tuned. The generated mode-locked single-soliton Kerr comb is robust and insensitive to perturbations. Even when the thermal effect induced by the absorption of the intracavity light is taken into account, the proposed single pulse trigger approach remains valid without requiring any thermal compensation means.**

## 1. Introduction

Since the use of self-referencing technique to stabilize carrier-envelope phase (CEP) was demonstrated in 2000 [1], optical frequency combs lead to revolutionary breakthrough in many fields such as spectroscopy, metrology, communications, biochemical sensing, etc., [2-4]. Kerr frequency combs, as a branch of the frequency comb family, can be generated through modulation instability (MI) and cascaded four-wave mixing (FWM) in externally driven nonlinear microresonators [5-7]. Microresonators, typically structured as disk [5,8], toroid [9,10], or ring [11,12], have the advantages of compactness, high finesse, and CMOS compatibility. Externally driven microresonators are intrinsically driven-damped nonlinear systems, which normally exhibit multistable and chaotic dynamics depending on the system parameters [13]. Such dynamical behaviors have been observed in the experiments of microresonator-based Kerr frequency comb formation [14]. Temporal dissipative cavity solitons, which are stable solitons superimposed on a continuous-wave (CW) background, can be achieved through comb formation. Especially, the single cavity soliton (SCS) state shows many attractive characteristics, e.g. the generation of few-cycle femtosecond pulses at microwave rates and octave spanning phase-locked comb spectra [14-17]. However, straightforward deterministic generation of the desired single-soliton state remains a challenge. The typical approach to obtain the single-soliton Kerr combs is to tune the pump frequency across a resonant frequency of the cavity from blue-detuned side to effectively red-detuned side [14-20]. Various improvements to the basic pump tuning scheme have been proposed. They include the combined forward and backward pump tuning to successively reduce the number of solitons generated in the microresonator [21,22], a two-step "power kicking" protocol to overcome the thermal destabilization effect [23,24], and a specified tuning pathway obtained by prior scanning of the parameter space to achieve the deterministic SCS by avoiding the chaotic and unstable states [25]. However, any pump tuning scheme requires tunable lasers which in general suffers from broad linewidth and thus high noise density to degrade the comb stability. Tuning the cavity resonance by current-controlling a resistive heater has been proposed to excite single-soliton with fixed-frequency pump lasers [26]. Recent works showed that electrically adjusting the free-carrier lifetime of a silicon microresonator can also lead to single-soliton mode-locking with a fixed-frequency pump [27]. Although these tuning based approaches show promising performances, it would be more attractive to straightforwardly and deterministically generate the SCS state without any active control. Phase modulation of the pump field has been proposed to excite the SCS instead of tuning approaches [28-32]. However, the roundtrip times of compact microresonators are typically in the order of a few picoseconds. Currently there are no commercially available phase modulators that could achieve such high-speed operation.

In this paper, we propose to generate single-soliton Kerr combs by seeding the CW pumped microcavity with a single shot pulsed trigger.

The proposed method does not require any frequency tuning process. The stable SCS is achieved by using a fixed frequency CW pump injected at a properly chosen frequency detuning from the resonance of cold cavity. Although similar cavity soliton excitation schemes by using addressing pulse had been proposed in spatial structures [33,34] and fiber cavities [35-37] for optical memory and trapping applications, to the best of our knowledge, such triggering scheme has never been proposed for Kerr comb generation in ultra-compact (typically several picoseconds per roundtrip) microresonators. In microresonators the cavity roundtrip time is typically either comparable to or even shorter than the duration of the trigger pulse. Thus the full trigger pulse is likely to cover multiple cavity roundtrip time. The coherent stacking of the trigger pulse inside the cavity may cause destructive perturbation to the intracavity field. Furthermore, the intrinsic strong thermal effects of microresonators due to the small volume and high absorption will unavoidably induce frequency shift of the cavity resonances which will then destabilize the cavity solitons. Neither problem exists in the long fiber cavities or large spatial cavities. Hence, the proposed trigger approach in microresonators is novel and unique.

## 2. Principles and methods

Figure 1(a) illustrates the proposed Kerr comb generation scheme. The combined field of a CW pump and a trigger pulse is injected into the microring resonator. The trigger pulse injects energy to those comb lines within the resonant profiles of the cavity and simultaneously covered by the spectrum of the trigger pulse. Most importantly, the phases of the comb lines are intrinsically synchronized as they are generated from the same single pulse trigger. Such coherently seeded comb lines accelerate the Kerr comb generation process and avoid the random build-up of pulses through modulation instability (MI).

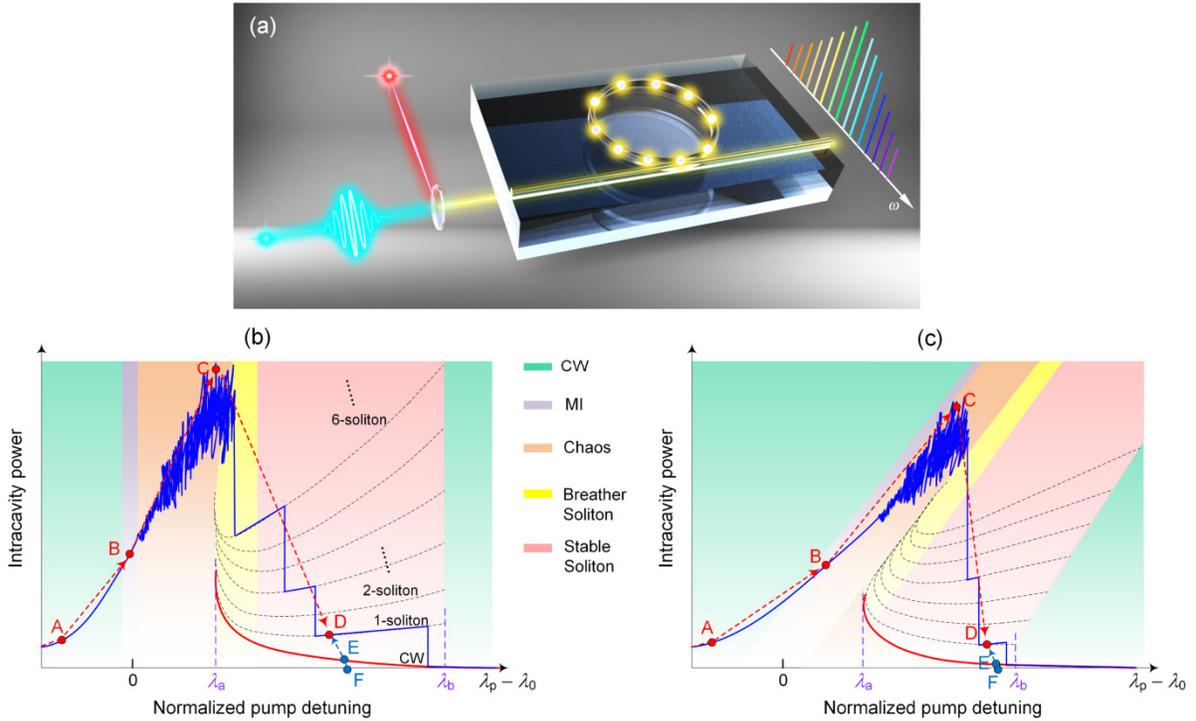

Fig. 1. (a) Schematic diagram of the proposed single-soliton Kerr comb generation. (b) and (c) are the illustrative diagrams of the variation of averaged intracavity power versus pump-resonance detuning in the Kerr comb excitation process when (b) only Kerr effect, and (c) both Kerr and thermal effects are considered. The blue solid curves show the possible intracavity power evolution to approach a single-soliton state in conventional pump frequency tuning scheme. The black dashed curves indicate the states with different number of intracavity solitons [14,21]. The lower red solid curve indicates the stable cold state with CW field only in the cavity. The evolution path to the single-soliton state utilizing the pump frequency tuning scheme (A→B→C→D) and the proposed trigger scheme (F→E→D) are indicated by the red and blue dashed arrows, respectively. Different colored regions indicate different intracavity states. MI stands for modulation instability. $\lambda_p$ is the pump wavelength, and $\lambda_0$ is the closest cold-cavity resonant wavelength. $\lambda_a$ and $\lambda_b$ mark the wavelength detune region where the SCS state exists as indicated in (b) and (c).

The wavelength of the CW pump is fixed and red-detuned [$\lambda_a < (\lambda_p - \lambda_0) < \lambda_b$] with respect to the closest cold-cavity resonant wavelength ($\lambda_0$), where $\lambda_p$ is the pump wavelength, $\lambda_a$ and $\lambda_b$ mark the wavelength detune range that SCS state exists, as indicated in Figs. 1(b) and 1(c). The evolution path to the SCS state in the proposed trigger scheme is rather different from that of the conventional pump tuning scheme. The schematic diagrams in Figs. 1(b) and 1(c) show the variation of the average intracavity power versus the pump-resonance detuning, which have terraced soliton solution states (black dashed curves). Kerr effect is included in both figures but thermal effect is considered in Fig. 1(c) only [14,21]. From Figs. 1(b) and 1(c), as the pump wavelength is increased from the blue-detuned side to the effectively red-detuned side, the cavity undergoes the (A) CW, (B) MI, (C) chaotic which is also called unstable MI, and (D) SCS state along the path A→B→C→D. In the proposed trigger scheme, the cavity starts from (F) cold state in the red-detuned side, first jumps to (E) a stable cold equilibrium CW state, and then to (D) the SCS state along the path F→E→D, by the combined action of the CW and trigger pulse. The inclusion of thermal effect tilts the response curve, as shown in Fig. 1(c), but the dominant behaviors shown in Figs. 1(b) and 1(c) are similar.

We note that the intracavity power of the SCS state (D) is much lower than that of the chaotic state (C). To reach the SCS state from the chaotic state, the cavity will have to be significantly cooled down. The excess energy should be quickly shed, otherwise the SCS state cannot be accessed and sustained. The difficulty to dissipate the excess energy is the reason that the resonance is typically lost when the pump wavelength enters the effectively red-detuned region in conventional pump tuning experiments. Thus special tuning and/or power kicking methods, which are equivalent to thermal annealing of the microcavity, is required to obtain and stabilize the SCS state from the chaotic state [14,23,24]. Despite the use of these special means, conventional pump tuning scheme to reach the SCS state remains unpredictable because it starts from a chaotic state. In the proposed trigger scheme, as the microcavity will not go through any chaotic state, the final state is deterministic. As the variation of intracavity energy is much smaller than that of the pump tuning scheme, the proposed trigger scheme is more robust to thermal perturbations. Since the trigger signal is the combination of a CW pump and a pulse, the light injected into the resonator is not a fixed quantity but varies significantly. Thus for the proposed trigger scheme, the mean-field Lugiato-Lefever equation (LLE) is no longer adequate to describe the system. We model the evolution of the optical field inside the cavity using the Ikeda map [38,39],

$$\psi^{(m+1)}(z=0,\tau) = \sqrt{\theta}\psi_{\text{in}}^{(m)}(\tau) + \sqrt{1-\theta}e^{-i\delta_0}\psi^{(m)}(z=L,\tau), \quad (1)$$

$$\frac{\partial \psi^{(m)}(z,\tau)}{\partial z} = -\frac{\alpha_0}{2}\psi^{(m)} + i\sum_{k\geq 2}\frac{i^k \beta_k}{k!}\frac{\partial^k \psi^{(m)}}{\partial \tau^k} + i\gamma(1+i\tau_s\frac{\partial}{\partial \tau})|\psi^{(m)}|^2 \psi^{(m)}. \quad (2)$$

where $\psi^{(m)}(z,\tau)$ is the slowly-varying envelope of the intracavity optical field in the $m$-th roundtrip, $\tau$ is the retarded time, $\alpha_0$ is the linear loss coefficient, $\theta$ is the coupling coefficient between the microring and the bus waveguide, $\beta_k$ is the $k$-th order dispersion coefficient, $\gamma$ is the nonlinear coefficient, $L = 2\pi r$ is the roundtrip length, $r$ is the radius of the microring, $\delta_0 = t_R(\omega_0 - \omega_p) = 2\pi l - \phi_0$ is the detuning between the CW pump frequency $\omega_p$ and its closest $l$-th order cold-cavity resonant frequency $\omega_0$, $\phi_0$ is the linear phase delay of the intracavity field in a single roundtrip, and $\tau_s = 1/\omega_p$ is the optical shock time constant. The dispersion of the nonlinearity is negligible in Si$_3$N$_4$ microring resonator [17,18,40,41]. The combined injection signal of a CW pump and a single shot pulse during the $m$-th roundtrip is given by

$$\psi_{\text{in}}^{(m)}(\tau) = \psi_{\text{cw}} + \sqrt{P_t}\exp\left[-i\Delta\Omega(\tau+mt_R)\right]\text{sech}\left[(\tau+mt_R-\delta t)/\tau_t\right], \quad (3)$$

where $\psi_{\text{cw}}$ is the CW pump field, $P_t$ and $\tau_t$ represent the peak power and duration of the trigger pulse, respectively. $\delta t$ represents the temporal central position of the trigger pulse. $\Delta\Omega = 2\pi\Delta f = 2\pi(f_t - f_{\text{cw}})$ is the central frequency offset of the trigger with respect to the CW pump, where $f_t$ and $f_{\text{cw}}$ are the central frequencies of the trigger pulse and the CW field, respectively. The fast variation of the combined injection signal is completely described by Eq. (3).

Thermal effects become significant only after the formation of the SCS but contribute little during the initial transient Kerr process. Therefore, we first study the trigger approach on SCS generation without including the thermal effect in Eqs. (1) and (2). The thermal effect is discussed and studied individually.

## 3. Results and discussion

### A. SCS excitation with a single shot pulse trigger

A single shot pulse trigger has the obvious advantages that none of the synchronization, timing jitter, and CEP stabilization problems will arise. A single shot pulse trigger can also be considered as a pulse train trigger with an extremely long period which is much longer than the build-up time of the SCS (typically several nanoseconds). If a single shot pulse trigger is sufficient to excite the SCS, we can switch off the trigger laser source after launching a single pulse. Thus any low repetition rate pico- or femto-second lasers with sufficient peak power and single pulse energy can be used to kick start the microresonator to reach the SCS state.

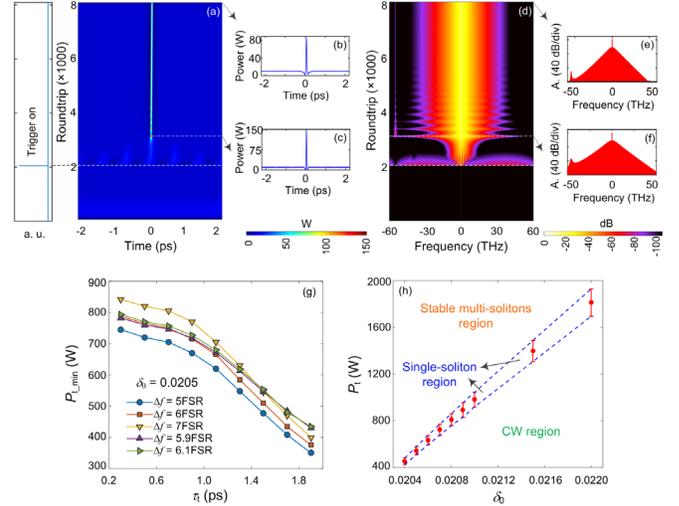

Fig. 2. (a)-(c) Temporal and (d)-(f) spectral evolutions of the intracavity field and the instantaneous profiles at the final and 3084-th roundtrips, respectively, when a single shot pulse trigger with a peak power of 520 W and an FWHM of 1.5 ps is injected. (g) $P_{t\_min}$ versus $\tau_t$ at different $\Delta f$. (h) Different operation regions within the parameter space of $P_t$ and $\delta_0$ with $\tau_t = 1.5$ ps and $\Delta f = 6\cdot$FSR. $\beta_2 = -59$ ps$^2$/km, $\gamma = 1$ W$^{-1}$/m, $\alpha_0 L = 0.012$.

To demonstrate the effectiveness of the proposed trigger approach, we use a Si$_3$N$_4$ microring resonator fabricated on silicon oxide substrate as an example, which is similar to that reported in [12]. The resonator has a free spectral range (FSR) of 226 GHz, a radius of $r = 100$ μm, and a coupling coefficient between the bus and ring waveguides of $\theta = 0.25\%$. The loaded Q-factor is ~1×10$^5$. The detuning $\delta_0$ is fixed to 0.0205 here. A single shot pulse with a peak power of 520 W is launched at the 2050-th roundtrip, accompanying the CW pump, as shown in the inset of Fig. 2(a). After the injection of the trigger pulse, several weak pulses appear in the cavity and eventually the central one evolves into a stable SCS. No chaotic or MI states are experienced. Fig. 2(b) shows the intracavity temporal waveform at the final roundtrip of the simulation. The SCS has a peak power of 86 W and FWHM of 43.4 fs. The waveform at the 3084-th roundtrip is shown in Fig. 2(c), which is captured at the first peak of a damped oscillation in the evolution. The cavity soliton has a higher peak power of 143 W and a smaller FWHM of 32 fs when compared with the final state. Figs. 2(d), 2(e), and 2(f) show the spectral evolution, spectral profiles of the intracavity field at the final and the 3084-th roundtrips, respectively. Fig. 2(d) shows that an initial broadband gain is introduced by the high energy single shot pulse trigger, which initiates and accelerates the generation of the Kerr comb lines.

Figure 2(g) shows the minimum $P_t$ required for SCS generation versus the pulse width $\tau_t$ for different central frequency offset $\Delta f$. The required $P_{t\_min}$ decreases when $\tau_t$ increases. For the $\Delta f$ values which are integer multiples of the FSR, a larger $\Delta f$ corresponds to a larger $P_{t\_min}$. The relationship is reasonable since larger $\Delta f$ will reduce the period of the beating oscillation caused by the interference between the trigger pulse and CW pump, which reduces the energy within each interfered peak. The reduction of energy has to be compensated at the expense of increasing $P_{t\_min}$. When $\Delta f$ deviates slightly from an integer multiple of the FSR, e.g. $\Delta f = 5.9\cdot$FSR and 6.1$\cdot$FSR, $P_{t\_min}$ deviates from the original curve especially when $\tau_t$ is larger than 1 ps. For such $\Delta f$, the roundtrip time will no longer be an integer multiple of the oscillation period, which may lead to destructive interference because of the coherent stacking of different parts of the pulse that leaked into adjacent roundtrips. Such perturbation will become significant when the trigger pulse is relatively long. For example, with a trigger pulse width of 1.5 ps, a $\Delta f$ deviation of ±0.1$\cdot$FSR from 6$\cdot$FSR will increase the power threshold by ~40 W. For the same deviation of $\Delta f$, the power penalty is only ~5 W if the trigger pulse width is 0.5 ps only. With a 0.5 ps pulse, the SCS state can be excited even when $\Delta f$ is decreased from 6.0$\cdot$FSR to

5.5·FSR, i.e. the central frequency of the pulse locates at the middle of two resonances of the microcavity, with a fixed peak power equal to $P_{t\_min}$ at $\Delta f = 6.0$·FSR. Thus, with an ultrashort trigger pulse whose spectra can cover several FSR, the SCS excitation is insensitive to the central frequency offset of the trigger pulse. In Fig. 2(g), the $P_{t\_min}$ curves of $\Delta f = 5.9$·FSR and 6.1·FSR begin to deviate from that of $\Delta f = 6.0$·FSR when $\tau_t$ is about 1 ps. Thus, the best choice of the trigger pulse width should be <1 ps for the configuration of the microresonator under studied. Fig. 2(h) shows the different working regions in the parameter space ($P_t$, $\delta_0$) at $\tau_t = 1.5$ ps and $\Delta f = 6$·FSR. For the $\delta_0$ considered, there is always a band of $P_t$ bounded by the two blue dashed curves that ensures SCS generation. The region above this band lead to stable multi-solitons states, while only CW states can be found in the region below this band. .

A single shot pulse trigger also makes it feasible to manipulate the number and temporal locations of the CSs generated by varying the peak power and temporal location of the trigger pulse. We believe such simple and robust SCS excitation scheme will be adopted for realizing ultrahigh-speed optical memory in monolithic microresonators.

### B. Deterministic SCS excitation

SCS excitation using the conventional frequency tuning approach is not deterministic because the intracavity field must undergo the chaotic state before the SCS state can be reached [18,25]. We compare the conventional pump frequency tuning approach with the proposed single shot pulse trigger approach by introducing temporal perturbations to the driving field. The perturbation is defined as $\sigma = \eta \tilde{N} \exp(i2\pi \tilde{U})$. Here $\tilde{N}$ is a normally distributed random variable with zero mean value and standard deviation of 1, $\tilde{U}$ is a uniformly distributed random variable between 0 and 1, and $\eta = 1\times 10^{-8}$ is the factor denoting the amplitude of the perturbation. The other parameters used in the pump tuning case are same to those in Fig. 2 except for the varied $\delta_0$.

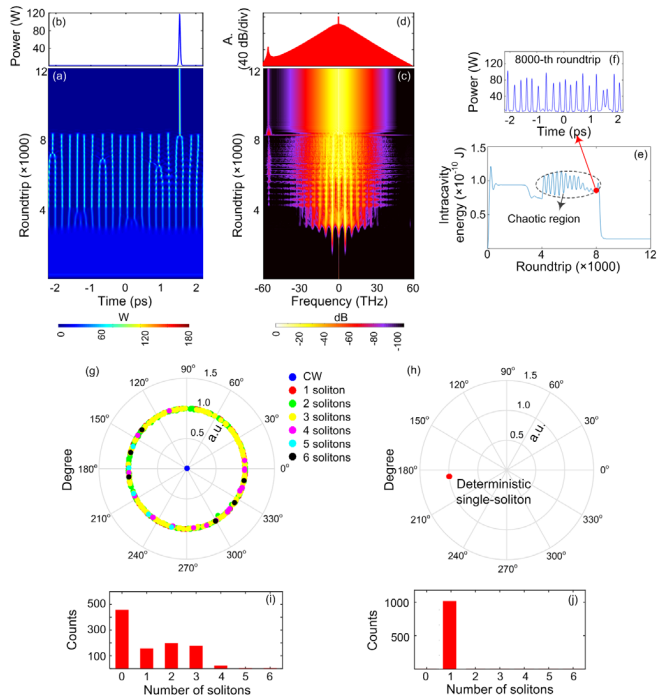

Fig. 3. (a)-(b) Temporal and (c)-(d) spectral evolutions of the intracavity field and the instantaneous profiles at the final roundtrip. (e) The intracavity energy at each roundtrip. (f) Instantaneous temporal profile at the 8000-th roundtrip of the pump frequency tuning approach. 1,000 simulation results using (g) the pump frequency tuning and (h) the proposed trigger approaches, respectively. The counts of different soliton states of the 1,000 simulations using (i) the pump frequency tuning and (j) the trigger approaches, respectively.

Figures 3(a) and 3(c) respectively show the temporal and spectral evolutions of the cavity field using the pump frequency tuning approach. Figs. 3(b) and 3(d) show the instantaneous temporal and spectral profiles of the cavity field at the final roundtrip (12,000-th) of the simulation, respectively. $\delta_0$ is initially set to –0.005, and increased to 0.018 and 0.035 at the 4000-th and 8000-th roundtrips, respectively. MI and chaotic states of the intracavity field are observed before the excitation of SCS. Multi-soliton states could also be excited if $\delta_0$ is chosen to a value between 0.018 and 0.035 at 8000-th roundtrip. Fig. 3(e) shows the evolution of the intracavity energy during the dynamics. The chaotic region is clearly indicated by the large oscillations of the energy. Fig. 3(f) shows the instantaneous temporal profile at the 8000-th roundtrip, which presents a group of pulses with random-like locations and amplitudes in the chaotic state. The results of 1,000 simulations for the pump frequency tuning and trigger approaches are shown in Figs. 3(g) and 3(h), respectively. The temporal positions and peak power of the solitons are plotted as the azimuthal angle and radial position in polar coordinates. The temporal positions in the range of [–$t_R$/2, $t_R$/2] are mapped into azimuthal angles in [0°, 360°], corresponding to the positions in the microring. The angular positions of all intracavity solitons are recorded despite of the final state. The peak power is normalized to the maximum peak power of the single-soliton case. It is clear that the final soliton states are not deterministic. The number of intracavity solitons varies from zero to six. Even CW states are observed. In contrast, the proposed trigger approach always deterministically excites the cavity to the SCS state despite the perturbations, as shown in Fig. 3(h). Figs. 3(i) and 3(j) show the counts of different soliton states with the pump tuning and trigger approaches, respectively. It is surprising that CW state without any intracavity soliton is the mostly observed state with the pump frequency tuning approach. The count of single-soliton with the pump frequency tuning approach is only 154 out of 1,000, while single soliton state is obtained in all of the 1,000 simulations with the proposed trigger approach.

### C. Thermally self-stabilized SCS

Compared with large volume fibers or spatial cavities, the thermal effect is much stronger for nanoscale microcavities [42]. The dissipated energy heats up the microcavities and induces changes in the refractive indices. Thus, the resonant frequencies of the cavities are shifted by the thermo-optic effect and the pump-resonance detuning cannot be modeled as a fixed value any more. The development of athermal optical materials and advanced cooling techniques are promising in overcoming the thermal problem, but up to now, these techniques still cannot fully avoid the thermal effect in microcavities [43].

The thermal effect leads to the thermal instability of the effectively red-detuned region [42], which forces the frequency scanning to start from the blue-detuned side in conventional pump tuning schemes. Thermal instability of the red-detuned region is the phenomenon that a pump initially injected at the near red-detuned side of a microresonator eventually operates in the effectively blue-detuned side because of the thermal effect. The observed thermal instability arises from the thermally-induced multistable nature of the microresonators. From Fig. 1(c), if the pump at $\lambda_p$ is initially injected at the red-detuned region $0 < (\lambda_p - \lambda_0) < \lambda_a$ of a cold microresonator, the combined thermal and Kerr effects will push the microcavity to operate in one of the states (CW or MI) on the blue solid curve above, which is in the effectively blue-detuned side [14,21,42]. If the pump starts the frequency scanning from a wavelength detune $(\lambda_p - \lambda_0) > \lambda_a$, the microcavity will stay at a cold CW state represented by the lower red solid curve in Fig. 1(c). As the cold CW state is very stable and has a lower intracavity power, it cannot reach the soliton states with higher intracavity power unless additional energy is injected into the microcavity. Scanning the pump frequency from the red-detuned region towards the blue-detuned side will eventually run into the thermal instability (when $(\lambda_p - \lambda_0) < \lambda_a$) and will not arrive at any of the soliton states.

In the proposed trigger scheme, the CW pump wavelength $\lambda_p$ is not tuned but fixed to a value in the region $\lambda_a < (\lambda_p - \lambda_0) < \lambda_b$ at a cold microresonator state (point F). Thus the thermal instability of red-detuned region is avoided. The microresonator will therefore first operate in the stable cold CW state (point E) and then get pushed up to the SCS state (point D) by the trigger pulse. We will show that the energy variation of this transition is small which only results in a small thermal shift, thus does not destroy the SCS state. We include the thermal effect into the model and divide the temporal triggering procedure of the SCS states into two stages, i.e. the CW and CS stages. In the CW stage, only the CW pump accumulates in the cavity and reaches the thermally stabilized state. Along the intracavity power accumulation, the resonant frequency of the cavity gets red-shifted towards the pump frequency, thus the pump-resonance detuning is reduced. Then, in the CS stage, the single shot trigger pulse is injected to excite the SCS state. If the thermally shifted pump-resonance detuning at the end of the CW stage still locates in the detuning region that supports CSs, then the SCS can be excited by the trigger pulse in the early CS stage. The thermal effect does not affect the excitation of SCS because the thermal response time of $Si_3N_4$ waveguide is typically microseconds or sub-microseconds [44,45], which is much longer than the excitation time of CS (~1000 roundtrips, 4 ns). The thermally shifted detuning is negligible in such short transient time. However, the increase of intracavity energy due to the injected pulse trigger does further shift the cavity resonant frequency, which might annihilate the excited SCS after thermal relaxation.

We first study the thermal effect in the CW stage. Since the intracavity CW field does not vary significantly between consecutive roundtrips and considering the extremely long thermal stabilization process, we use a modified LLE to improve the computation efficiency,

$$t_R \frac{\partial \psi}{\partial t} = \sqrt{\theta}\psi_{in} - \frac{\alpha_0 L + \theta}{2}\psi - i(\delta_0 + \delta_{therm})\psi$$
$$+ iL\sum_{k\geq 2}\frac{i^k \beta_k}{k!}\frac{\partial^k \psi}{\partial \tau^k} + i\gamma L(1 + i\tau_s \frac{\partial}{\partial \tau})|\psi|^2 \psi, \quad (4)$$

$$\frac{d\delta_{therm}}{dt} = -\frac{\delta_{therm}}{\tau_0} + \frac{\xi}{t_R}\int_0^{t_R}|\psi|^2 dt, \quad (5)$$

where $\delta_{therm} = t_R(\omega_{therm} - \omega_0)$ is the thermally induced detuning shift, $\omega_{therm}$ is the thermally shifted cavity resonant frequency. $\tau_0$ is the thermal response time determined by the material of the microresonator and the thermal dissipation in experiments. $\xi$ is the coefficient representing the detuning shift in response to the average intracavity power. We use $\tau_0$ = 100 ns and $\xi$ = –4.5×10$^4$ W$^{-1}$s$^{-1}$ in the simulation according to a practically fabricated high-Q (~1.1×10$^6$) $Si_3N_4$ microresonator [45,46], which has comparable volume and parameters to the one used in Fig. 2 except a smaller propagation loss of $\alpha_0 L$ = 0.0012. Comparing with the thermo-optic effect of $Si_3N_4$, the contribution of thermal expansion effect is much smaller and can be neglected [44,45]. Hence, $\xi \propto \wp C \alpha_0 L$ depends on the thermo-optic coefficient $\wp$ of $Si_3N_4$, the heat capacity $C$ of the microring, and the ratio that intracavity power is absorbed and converted to heat. We can use the $\xi$ value at $\alpha_0 L$ = 0.0012 as a reference to calculate $\xi$ at other loss parameters.

Figure 4(a) shows the stabilized detuning $\delta_1 \equiv \delta_0 + \delta_{therm}$ at the end of CW stage versus initial detuning $\delta_0$ under different propagation loss and pump power. Simulation is carried out for 5×10$^6$ roundtrips (~22 μs) to guarantee thermal stabilization in the CW stage. We find that, for a given pump power and loss, there is a lowest initial detuning $\delta_{0\_min}$ to ensure that the shifted detuning will still remain within the red-detuned side. The final stabilized detuning $\delta_1$ will be positive (effectively red-detuned) only if $\delta_0 \geq \delta_{0\_min}$, which is a necessary condition for CS excitation. When $\delta_0 < \delta_{0\_min}$, $\delta_1$ becomes negative (effectively blue-detuned), thus no CS can be excited. The results are consistent with that shown in [42], which indicates two kinds of thermal equilibriums, i.e. stable warm equilibrium (effectively blue-detuned) and stable cold equilibrium (effectively red-detuned) CW states can be reached. With the increase of propagation loss, the threshold $\delta_{0\_min}$ also increases since more intracavity power is converted to heat. $\delta_{0\_min}$ will also increase when the pump power $P_{cw}$ increases. The detuning value $\delta_{0\_min}$ gives the critical value to reach an effectively red-detuned state, which corresponds to the lowest red solid curve in Fig. 1(c).

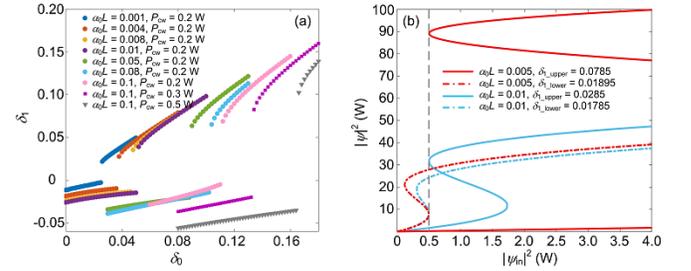

Fig. 4. (a) Stabilized $\delta_1$ at the end of CW stage starting from a cold cavity as a function of initial detuning $\delta_0$ under different propagation loss and pump power. Other parameters: $\gamma$ = 1.4 W$^{-1}$/m, $L$ = 628 μm, $\theta$ = 0.0025, $t_R$ = 4.425 ps. (b) S-curves with the border detuning values $\delta_{1\_lower}$ and $\delta_{1\_upper}$ for the pump power of 0.5 W when $\alpha_0 L$ are 0.005 and 0.01, respectively.

It is known that the CS states exist almost in the same bistable region of the stationary CW solution solved by Ikeda map or LLE [16,35,41]. The two models give the same bistable region when the nonlinear length is much longer than the cavity length [17,41]. Since the thermal effect responds in a long time scale, we can focus on the initial stage of the CS stage and obtain the stationary CW solution by setting the derivative terms in Eq. (4) to zeros and neglecting the variation of $\delta_1$ in the transient triggering process. The equation becomes

$$Y = \frac{\gamma^2 L^2}{\theta}X^3 - \frac{2\delta_1 \gamma L}{\theta}X^2 + \frac{(\delta_1^2 + \alpha^2)}{\theta}X, \quad (6)$$

where $Y = |\psi_{in}|^2$ is the pump power, $X = |\psi|^2$ is the intracavity power, $\alpha = (\alpha_0 L + \theta)/2$ is the total loss per roundtrip. The turning points of the function $Y(X)$ can be calculated by setting the first derivative $dY/dX$ to zero as

$$3\gamma^2 L^2 X^2 - 4\delta_1 \gamma L X + (\delta_1^2 + \alpha^2) = 0. \quad (7)$$

The function $X(Y)$ given by Eq. (6) has a bistable region when Eq. (7) has two different real roots of $X$, which requires $\delta_1^2 > 3\alpha^2$. The two real roots corresponding to the two turning points of the bistable curve are given by

$$X_{1,2} = \frac{2\delta_1 \pm \sqrt{\delta_1^2 - 3\alpha^2}}{3\gamma L}. \quad (8)$$

Substituting these roots into Eq. (6), we obtain the minimum and maximum of $P_{cw}$ in the bistable region for given values of $\delta_1$ and $\alpha$. By sweeping the values of $\delta_1$, the limited bistable regions are found in the 2-dimensional ($P_{cw}$, $\delta_1$) space with $\alpha_0 L$ values varying from 0.0012 to 0.012, shown in Fig. 5 as the green regions confined by the two orange boundary curves in each map. The two specific loss values, i.e. $\alpha_0 L$ = 0.0012 and 0.012, correspond to two practically fabricated $Si_3N_4$ microresonators [18,45,46]. The other two loss values of $\alpha_0 L$ = 0.004 and 0.008 are arbitrarily chosen between the two specific values. In each green bistable region, there are two boundary detuning values $\delta_{1\_upper}$ and $\delta_{1\_lower}$ of $\delta_1$ for each $P_{cw}$. In the detuning range of $\delta_{1\_lower} < \delta_1 < \delta_{1\_upper}$ with a given pump power $P_{cw}$, the system always has two stable solutions. Fig. 4(b) shows two sets of S-curves with the boundary detuning values $\delta_{1\_lower}$ and $\delta_{1\_upper}$ for a pump power $P_{cw}$ = 0.5 W. When the loss $\alpha_0 L$ is 0.005, the $\delta_{1\_upper}$ (red solid curve) and $\delta_{1\_lower}$ (red dashed curve) are calculated to be 0.0785 and 0.01895, respectively. The difference between $\delta_{1\_upper}$ and $\delta_{1\_lower}$ is 0.0595. When $\alpha_0 L$ is increased to 0.01, $\delta_{1\_upper}$ = 0.0285 (blue solid curve) and $\delta_{1\_lower}$ = 0.01785 (blue dashed curve) are obtained. Compared to that of $\alpha_0 L$ = 0.005, the difference between $\delta_{1\_upper}$ and $\delta_{1\_lower}$ is reduced to

0.0107, which means the range of valid $\delta_1$ becomes narrower. The trend of the variation agrees with that of the green regions in Fig. 5.

If the thermally-stabilized detuning $\delta_1$ falls into the region bounded by $\delta_{1\_upper}$ and $\delta_{1\_lower}$ at the end of the CW stage, SCS can be excited very quickly after the injection of the single shot trigger pulse. Fig. 5 shows the minimum possible values of $\delta_1$ (blue dots) that can be stabilized in the red-detuned side with their fitting (blue dashed curves). In experiments, the combinations of $P_{cw}$ and corresponding $\delta_1$ above the blue curves can be obtained by referring to the results in Fig. 4(a). Thus we can easily identify the valid regions (hatched regions in Fig. 5) for SCS excitation when thermal effect is taken into account, which are the overlapping area of the green and required $\delta_1$ regions. When the loss $\alpha_0 L$ is increased, i.e. the Q-factor of the microresonator is decreased, Figs. 5(a)-5(d) show that the region bounded by $\delta_{1\_upper}$ and $\delta_{1\_lower}$ becomes narrower. Meanwhile, the required minimum pump power $P_{cw}$ and minimum detuning $\delta_1$ in the hatched region are both increased. Since injecting a CW light with several watts of power into a nanoscale waveguide is not practical, it is hard to excite the SCS in low-Q $Si_3N_4$ microresonators under the influence of thermal effect. The threshold of the Q-factor defined as "low-Q" depends on the operating range of the pump power. For the given pump power ranging from 0 to 1.2 W, Fig. 5(d) shows that there is no valid region for a cavity with $\alpha_0 L = 0.012$ (Q-factor ~1×10$^5$). In contrast, when $\alpha_0 L$ is 0.0012 (Q-factor ~1.1×10$^6$), Fig. 5(a) shows that the hatched region is quite large, which means the selection of initial $\delta_0$ becomes very flexible.

From Fig. 5(a), we arbitrarily choose a pump power value of $P_{cw} = 0.03$ W and an initial detuning of $\delta_0 = 0.018$ which corresponds to a $\delta_1$ value within the hatched region, and simulate both the CW and CS stages with the inclusion of thermo-optics effect. Considering both the coherent stacking of the leaked pulse tails of the full trigger pulse and the fast variation of the intracavity field between consecutive roundtrips, we simulate the SCS triggering process in the CS stage using a modified Ikeda map, which is similar to that of Eqs. (1)-(3) except that we replace $\delta_0$ by $\delta_0 + \delta_{therm}$ and add the thermal dynamics governed by Eq. (5). We also assume that the variation of $\delta_{therm}$ is small in one roundtrip and update the $\delta_{therm}$ value in steps of $t_R$.

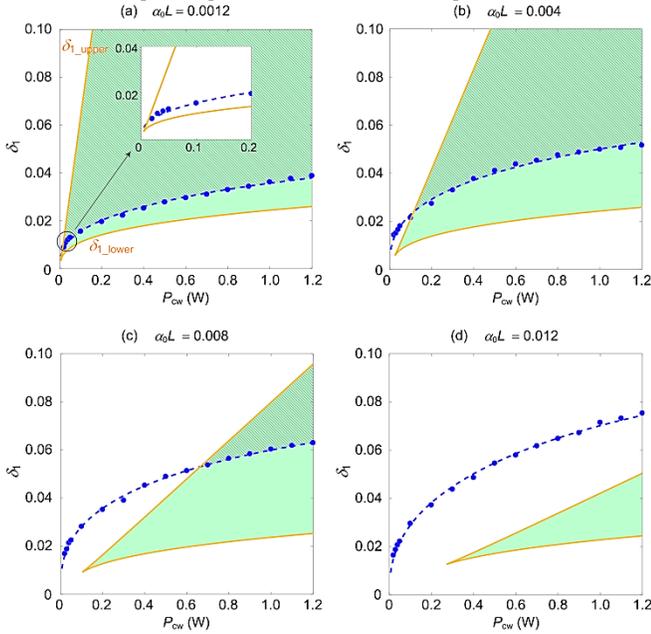

Fig. 5. The green regions are the possible CS excitation regions (bistable regions) bounded by the detuning $\delta_{1\_upper}$ and $\delta_{1\_lower}$. Blue dashed curves with dots show the minimum $\delta_1$ that can be thermally stabilized in the red-detuned side. Hatched regions indicate the valid regions for SCS excitation with thermal effect. The loss $\alpha_0 L$ is (a) 0.0012, (b) 0.004, (c) 0.008, and (d) 0.012, respectively. The inset in (a) shows the zoom-in view of the pump power ranging from 0 to 0.2 W.

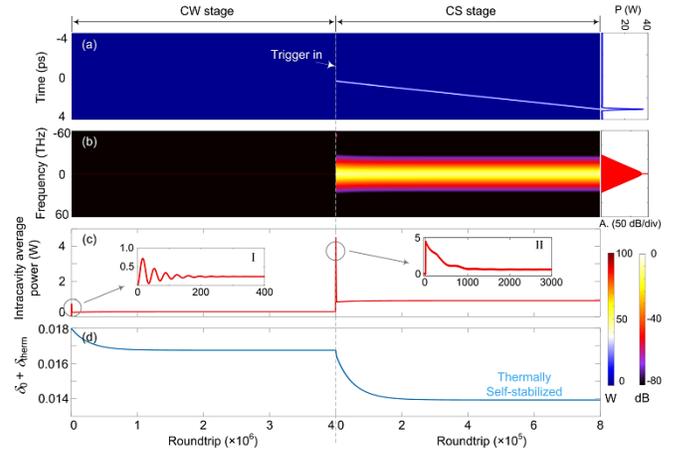

Fig. 6. (a) Temporal evolution and final instantaneous temporal profiles (b) spectral evolution and final instantaneous spectral profiles in the CW and CS stages. (c) Evolution of intracavity average power. Inset I and II show the zoom-in views in [0, 400] roundtrips of the CW stage and [0, 3000] roundtrips of the CS stage, respectively. (d) Evolution of $\delta_0 + \delta_{therm}$.

Figures 6(a) and 6(b) show the intracavity temporal and spectral evolutions along with the instantaneous temporal and spectral profiles at the end of the CS stage, respectively. The corresponding variations of intracavity average power and $\delta_0 + \delta_{therm}$ are shown in Figs. 6(c) and 6(d), respectively. The intracavity average power experiences a relaxation oscillation in the initial several hundreds of roundtrips (inset I of Fig. 6(c)), then increases slowly in the long thermal stabilization process and eventually converges to a fixed power of 0.2772 W. The detuning $\delta_0 + \delta_{therm}$ varies significantly in the first 1×10$^6$ roundtrips of the CW stage and slowly converges to 0.016753. The complete thermal stabilization of the cavity is achieved after ~2×10$^6$ roundtrips. A single shot 0.5 ps hyperbolic secant pulse with a peak power of 7.5 kW is launched at the end of CW stage to initiate the CS stage. The SCS is excited in ~1000 roundtrips (4.42 ns), and more importantly the newly excited SCS survives and is thermally self-stabilized in the subsequent evolution. The average power and $\delta_0 + \delta_{therm}$ are stabilized to 0.90375 W and 0.013933, respectively, in ~6×10$^5$ roundtrips of the CS stage. The stability of the SCS under thermal perturbation is the main advantage of the proposed pulse trigger approach. Unlike the conventional frequency tuning method, in which the tuning process is typically at microseconds level (comparable to or even longer than the thermal response time) and the intracavity field typically experiences the higher energy chaotic states, the SCS can be directly excited within several nanoseconds without going through any chaotic dynamics in the proposed pulse trigger approach. Hence, the thermal perturbation during the SCS excitation process is negligible as heating is a slow process. In the subsequent evolution at the thermal response time scale, the variation of the intracavity energy caused by the trigger pulse will further affect the detuning until the thermal equilibrium is reached. Since the energy variation is small, the transition from the CW state to the SCS state only results in a small thermal shift, which will not destroy the excited soliton state. We find that the final detuning $\delta_0 + \delta_{therm} = 0.013933$ is still in the green region and larger than the boundary value $\delta_{1\_lower} = 0.0073$. Thus, the SCS will not be annihilated, but slowly adjust the working point with respect to the gradually changed detuning by thermal effect, and eventually gets thermally self-stabilized. It should be pointed out that it is not recommended to excite the SCS with $\delta_1$ close to the lower boundary $\delta_{1\_lower}$, where breather CS states may appear [21,47,48]. The energy of the breather CSs is continually changing so that it is hard to be thermally self-stabilized. The exact thermally self-stabilization region of the SCS is not given in this paper, but will be studied in future work. However, we expect that this region will become broader for high-Q microresonators since the optical energy converted to heat is less, and the CSs valid region bounded by $\delta_{1\_upper}$ and $\delta_{1\_lower}$ will become broader too.

## 4. Conclusions

In conclusion, we propose to directly and deterministically generate single-soliton Kerr comb in a continuous-wave pumped microring resonator by seeding it with a pulsed trigger. Single cavity soliton can be straightforwardly and deterministically triggered by an energetic single shot pule without going through any multi-stable or chaotic states in the nonlinear microcavity. The feasibility of single shot trigger approach makes it possible to manipulate the number of solitons and their temporal locations inside the cavity by simply controlling the power and temporal location of the single pulse trigger. The proposed trigger approach can be applied to different type of microresonators as long as the microresonators do support cavity solitons. Moreover, we find that even under strong thermal effect the proposed trigger approach is valid to excite single cavity solitons in practically fabricated $Si_3N_4$ microresonators. No additional complex engineering is required for thermal compensation. Benefiting from the fast triggering process and relatively slow thermal dynamics caused by the intracavity optical heating, single cavity solitons can be easily excited within a broad parameters range and thermally self-stabilized in long-term evolutions. Since the triggering process is insensitive to the central frequency offset of the single shot trigger pulse when the pulse width is much shorter than the roundtrip time, the choice of trigger source is very flexible and feasible. The thermally self-stabilized single cavity solitons suggested that the proposed trigger approach can be applied to microresonators fabricated by current fabrication foundries to generate deterministic cavity solitons, which will pave the way to many applications of single-soliton Kerr combs.

## 5. Acknowlegement

Hong Kong Research Grants Council (PolyU152471/16E); The Hong Kong Polytechnic University (1-ZVGB); National Natural Science Foundation of China (61307109, 61475023, and 61475131); Shenzhen Science and Technology Innovation Commission (JCYJ20160331141313917); Beijing Youth Top-notch Talent Support Program (2015000026833ZK08).